\documentclass{article}
\usepackage{spconf,amsmath,graphicx}
\usepackage{subfigure}
\usepackage{times}
\usepackage{epsfig}
\usepackage{graphicx}
\usepackage{amsmath}
\usepackage{amssymb}
\usepackage{epstopdf}
\usepackage{cite}
\usepackage{subfigure}
\usepackage{enumerate}
\usepackage{caption2}
\usepackage{verbatim}
\usepackage{xcolor}
\usepackage{hyperref}
\usepackage{subfigure}
\title{Neural Video Compression using Spatio-temporal Priors}
\name{Haojie Liu, Tong Chen, Ming Lu, Qiu Shen, and Zhan Ma\thanks{ Emails: \{haojie, tong, ming\}@smail.nju.edu.cn, \{shenqiu, mazhan\}@nju.edu.cn}}

\address{School of Electronic Science and Engineering, Nanjing University}
\begin{document}
%\ninept
%
\maketitle
\begin{abstract}
The pursuit of higher compression efficiency continuously drives the advances of video coding technologies.  Fundamentally, we wish to find better ``predictions'' or ``priors'' that are reconstructed previously to remove the signal dependency efficiently and to accurately model the signal distribution for entropy coding. In this work, we propose a neural video compression framework, leveraging the spatial and temporal priors, independently and jointly to exploit the correlations in intra texture, optical flow based temporal motion and residuals. Spatial priors are generated using downscaled low-resolution features, while temporal priors (from previous reference frames and residuals) are captured using a convolutional neural network based long-short term memory (ConvLSTM) structure in a temporal recurrent fashion. All of these parts are connected and trained jointly towards the optimal rate-distortion performance. Compared with the High-Efficiency Video Coding (HEVC) Main Profile (MP), our method has demonstrated averaged 38\% Bjontegaard-Delta Rate (BD-Rate)  improvement  using standard common test sequences, where the distortion is  multi-scale structural similarity (MS-SSIM).
\end{abstract}

\section{Introduction}
\label{sec:intro}

Over the past three decades, successful video compression technologies have been following the similar {\it hybrid block-based transform and motion-compensation} framework with handcrafted coding tools, such as recursive block-size, directional intra prediction, discrete cosine transform (DCT), interpolation, context-adaptive entropy coding, etc, resulting in several well-known international standards, e.g., HEVC~\cite{HEVC} and emerging versatile video coding (VVC)~\cite{VVC_spec}. All of these and other technical explorations in video compression are trying to exploit and remove signal redundancy using ``causal priors'', e.g, reconstructed neighbor pixels, previous frames, context probability of neighbors, in video content, spatially, temporally and statistically~\cite{Gary_VideoCompressionBasics}, for better compact representation at the same quality.

Motivated by the recent advances in deep learning, a variety of deep neural network (DNN) based image/video compression methods were developed via end-to-end learned (not handcrafted) coding tools~\cite{balle2018variational,Liu_2018_CVPR_Workshops, rippel2017real,mentzer2018conditional,chen2017deepcoder,lu2018dvc, rippel2018learned}. Either conventional or recent emerging learning based video compressions are primarily trying to exploit the correlations between existing priors and ``pixels-to-be-coded''. In addition to previously reconstructed pixels and context probabilities used in traditional video coding methods,  DNN solutions could also generate hyperpriors in feature domain for better prediction.

\begin{figure*}[t]
     \centering
     \subfigure[]{\includegraphics[scale=0.36]{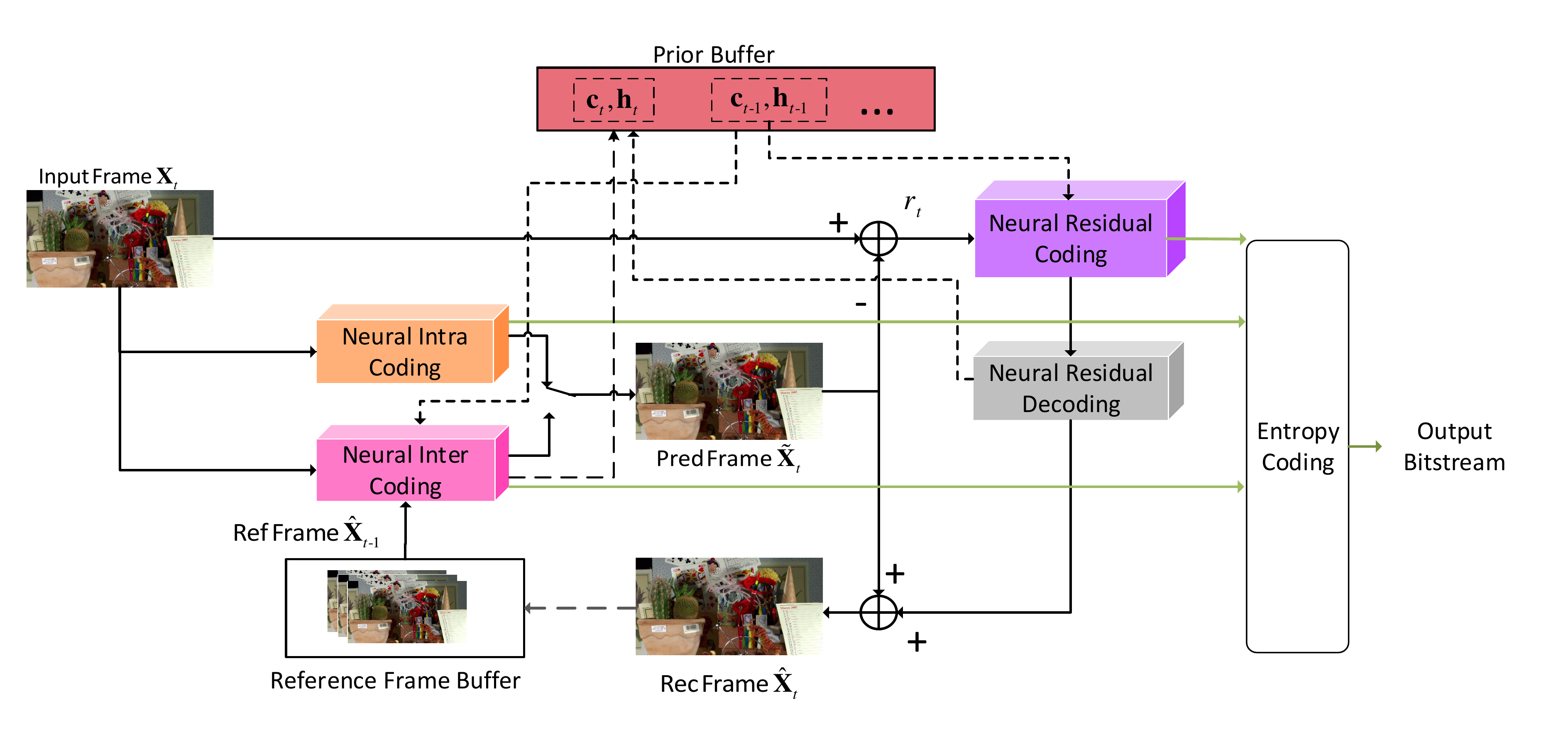} \label{sfig:overall}}\\
     \subfigure[]{\includegraphics[scale=0.34]{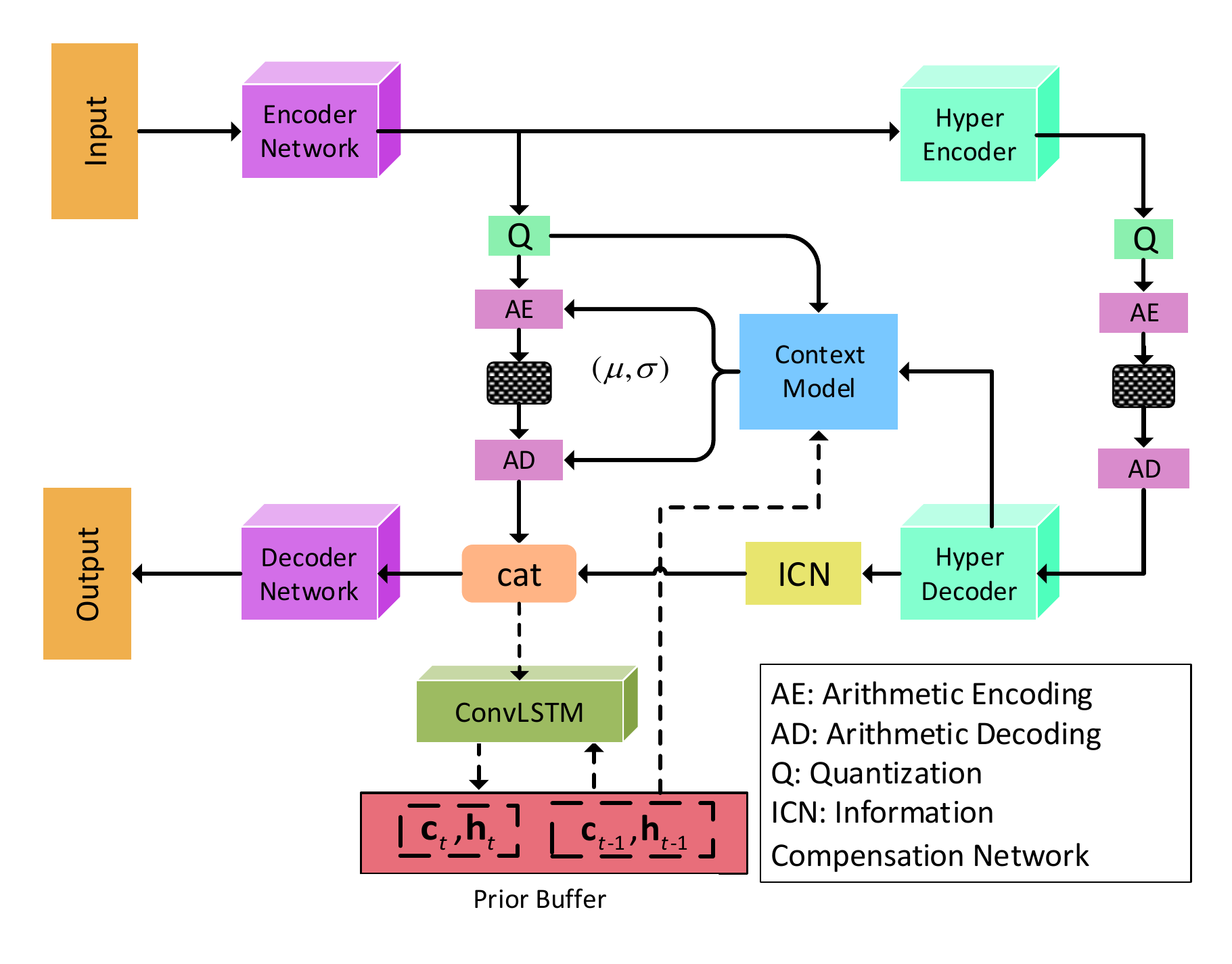} \label{sfig:intra_residual}}
     \subfigure[]{\includegraphics[scale=0.34]{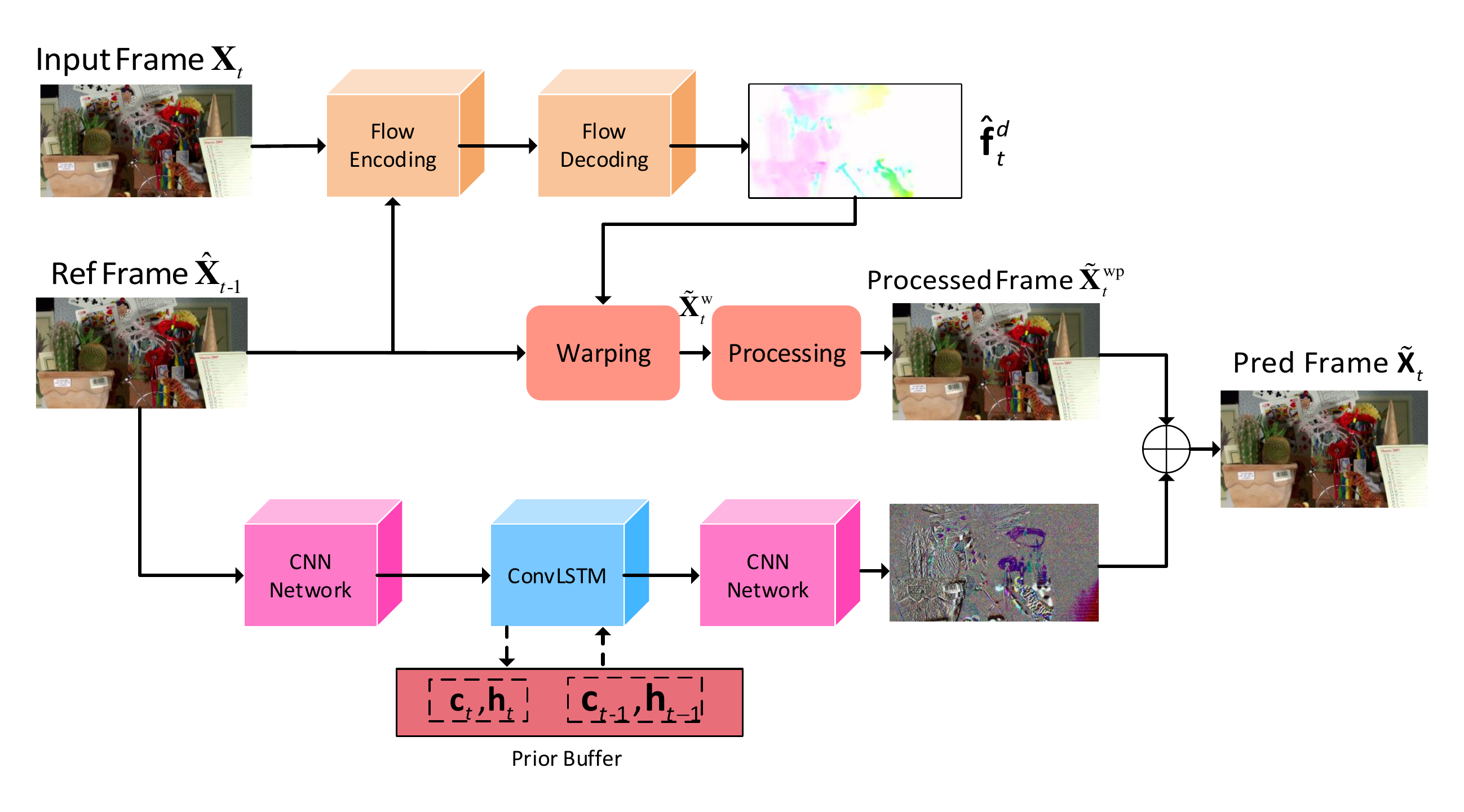} \label{sfig:inter_coding}}
     \caption{Neural video compression framework: (a) overall architecture with neural intra, inter and residual coding; (b)  intra and residual coding; (c) inter coding with flow processing and temporal recurrent update.}
     \label{fig:arch}
\end{figure*}

{In this work, we have presented a {\it neural video compression} (NVC) framework in Fig.~\ref{fig:arch}, leveraging the end-to-end learning to generate latent features for compact representation of spatial intra texture, temporal motion and statistical context probability. Joint spatio-temporal priors have been used extensively to improve the compression efficiency, for example, 1) spatial priors for both conditional probability modeling and reconstruction of intra texture (cf. Fig.~\ref{sfig:intra_residual});  2) temporal priors for frame reconstruction (cf. Fig.~\ref{sfig:inter_coding}); and 3) joint spatio-temporal priors for temporal predictive residual encoding (cf. Fig.~\ref{sfig:intra_residual}) (e.g., context probability) and reconstruction.}

{Spatial priors are generated using the low-resolution (e.g., via aggregated downscaling) representations from the same image content, while temporal priors are provided using the ConvLSTM~\cite{gers2001long,xingjian2015convolutional} to capture the long-short dependency of previously processed frames.
Temporal motion representation often plays an important role for video compression. Traditional methods adopt straightforward but effective variable block size based motion estimation to exploit the temporal correlations. But, in this work, we have turned to more fundamental {\it optical flow} for motion description instead.}

%\begin{figure}[t]
%     \centering
%     \includegraphics[scale=0.42]{intra_residual1.pdf}
%     \caption{Illustration of our intra coding and residual coding network. For residual coding, we add a ConvLSTM as temporal information updating in the dotted.}
%     \label{fig:intra_residual}
%\end{figure}

%\begin{figure}[t]
%     \centering
%     \includegraphics[scale=0.35]{inter.pdf}
%     \caption{Illustration of temporal recurrent network. $x_{wp}$ is the warped frame after processing. $\hat{x}_{t-1}$ denotes the former frame. It is useful for further enhancement of the reconstructed frame by frame mapping.}
%     \label{fig:inter}
%\end{figure}

We have evaluated our NVC for a low-delay, e.g., IPPP, coding structure, where except the first frame is encoded as an intra frame, all the rest frames are inter-coded with unidirectional forward prediction\footnote{Bidirectional prediction is deferred as our future study.}. Performance comparisons have been carried out with well-known H.264/AVC High Profile (HP)~\cite{H264AVC} and HEVC MP~\cite{HEVC}, using industry leading x264 (\url{https://git.videolan.org/git/x264.git}) and x265 (\url{http://x265.org/}). For fairness, both x264 and x265 are constrained with IPPP low-delay encoding configuration with the other parameters remained as default.  Among those standard common test video sequences, our NVC has demonstrated superior coding efficiency over both H.264/AVC HP and HEVC MP, e.g., $\approx$38\% BD-Rate~\cite{bjontegaard2001calculation} improvements against the HEVC MP. Note that distortion measurement used in evaluation is the MS-SSIM~\cite{wang2003multiscale} presented in decibel (dB) scale.

%IPPP coding mode which outperforms the well known H.264 and HEVC in terms of MS-SSIM.
%Our framework allows more efficient temporal information passing and a better context model based on multi-priors fusion. We have not introduced any refined inter RDO and showed our superiority as a baseline that different modules can be easily added into our architecture to improve the performance.
{}

\section{Neural Video Compression: From Model-Driven to Data-Driven Solution}
\label{sec:nvc}

Our NVC has attempted to define a way for efficient video compression through data-driven learning, rather than the traditional model-driven coding tool (e.g., transform model, motion model, etc) development. Details will be unfolded in subsequent paragraphs.

%Technical details will be unfolded for our NVC in subsequent

%In this section, we provide more details about each individual module such as intra coding, inter coding and residual coding.

\subsection{Neural Intra Coding}
Neural intra (texture) coding has tried to exploit the correlation within current video frame.
As shown in Fig.~\ref{sfig:intra_residual}, we utilize a variational autoencoder with embedded hyperpriors for high-efficiency image coding. Similar architecture is also used in~\cite{balle2018variational}.  We use deep residual learning (ResNet)~\cite{he2016deep} with generalized divisive normalization (GDN) transform based activation, but not conventional rectified linear unit (ReLU), for fast convergence in training and more compact latent features, in both encoder network $\mathbb E$ and decoder network $\mathbb D$. Convolution with parametric ReLU is used for hyper encoder and decoder. Quantization $Q$ is approximated by adding uniform noise in training, but carried out with {\tt ROUND}($\cdot$) operation in inference.

%\textcolor{red}{need some comments on probablity model to estimate rate and to encode}
An accurate probability distribution model of quantized features is the key for high-efficiency compression, not only for the arithmetic encoding (AE), but also for the rate estimation of rate-distortion optimization~\cite{Gary_RDO} and bit allocation.  We jointly leverage the autoregressive information (i.e., distribution of quantized features) and hyperpriors (i.e., distribution of decoded hyper-features) for conditional context probability $P$ modeling with accurate mean ($\mu$)  and variance ($\sigma$) prediction assuming a Gaussian distributed probability mass function, i.e.,
\begin{align}
  (\mu,\sigma) = {\prod_i} p\left(\hat{x}_i|\hat{x}_1,..., \hat{x}_{i-1},{\hat{\bf z}_t}\right).
  \label{eq:intra_feature}
\end{align}
Here,  $\hat{x}_1, {\hat {x}}_2,..., \hat{x}_{i-1}$ denote the causal (and possibly reconstructed) pixels prior to current pixel $\hat{x}_i$ and $\hat{\bf z}_t$ are the hyperpriors, for image $\hat{\bf X}_t$. Probability of each pixel symbol can be simply derived using
\begin{align}
  p_{\hat{x}|\hat{x}_1,..., \hat{x}_{i-1},\hat{\bf z}_t}&(\hat{x}|\hat{x}_1,..., \hat{x}_{i-1},\hat{\bf z}_t) \nonumber \\
  &\mbox{~~~~}  = {\prod_i} (\mathcal{N(\mu,\sigma)} *\mathcal{U}(-\frac{1}{2},\frac{1}{2})) (\hat{x}_i).
  \label{eq:intra_texture_dist}
 \end{align}

In addition to use hyperpriors $\hat{\bf z}_t$ for conditional probability improvement, we have developed an Information Compensation Network (ICN) to fuse and concatenate ({\bf cat} operation in Fig.~\ref{sfig:intra_residual}) decoded hyperpriors with latent features for better reconstruction.

%\textcolor{red}{some comments on Information Compensation Network}
%Both of our intra and residual coding network as are based on variational autoencoder(VAE) alike struture as shown in Fig~\ref{intra_residual}. Residual connections are added in generalized divisive normalization(GDN) transform to make more effecient and faster convergence in $E$ and $D$.

%During training, we add uniform noise to analog approximate quantization $Q$ and replace it with Round(.) function in actual use.

%In addition, information compensation network(I) is introduced to fully explore the hyper information as both probability estimation and final reconstruction.

%For our intra coding network, we just utilize the jointly autogressive information and hyperprior for entropy modeling. Here, we propose a 3D masked convolution based context model $P$ with priors fusion to predict the mean and variance for a Gaussian distribution to predict the actual probability using Eq.~\ref{Eq1}
%\begin{equation}
%  \mu,\sigma = {\prod_i} p(x_i|x_1,...,x_{i-1},z_h),
%  \label{Eq1}
%\end{equation}
%Here, $\mu$,$\sigma$ is a predicted matrix for a Gaussian distribution to provide the actual probablity. $x_1,x_2,...,x_{i-1}$ denotes the pixels before the current pixel and $z_h$ denotes the hyperprior.

\subsection{Neural Inter Coding} \label{ssec:inter_coding}
The key issues for improving the efficiency of temporal inter coding are two folds. One is to accurately represent the motion between consecutive frames, and the other is to have high-fidelity reconstructions for compensation.

%Our inter prediction network is composed of flow coding network, processing network and temporal recurrent network as shown in Fig.~\ref{inter}.

First, we use optical flow $f_t$ for accurate motion representation that are learned between consecutive frames, e.g., ${\bf X}_t$ and $\hat{\bf X}_{t-1}$, as shown in Fig.~\ref{sfig:inter_coding}. A compressed flow representation $\hat{f}_t$ is encoded into the bitstream for delivery.  For compensation, $\hat{f}_t$ is then decoded into $\hat{f}^d_t$ for warping with reference frame to have ${\bf X}^w_t$, i.e.,
\begin{align}
  \hat{f}^d_t &= {\mathbb F}_d\left({\mathbb F}_e\left(\hat{\bf X}_{t-1},{\bf X}_t\right)\right), \label{eq:decoded_flow}\\
  {\hat{\bf X}}^w_t &= {\bf warp}\left(\hat{\bf X}_{t-1},\hat{f}^d_t \right). \label{eq:warped_frame}
\end{align}
Here ${\mathbb F}_e$ and ${\mathbb F}_d$ represent the cascaded optical flow encoder and decoder network~\cite{ilg2017flownet}.
To avoid quantization induced motion noise, our flow network is first pre-trained with uncompressed frames ${\bf X}_{t-1}$ and ${\bf X}_t$. Then we replace the former one ${\bf X}_{t-1}$  using the decoded reference frame ${\hat{\bf X}}_{t-1}$ as described in Eq.~\eqref{eq:decoded_flow} and \eqref{eq:warped_frame}. Note that we have directly utilized the decoded  flow $\hat{f}^d_t$ for end-to-end training.

Oftentimes, warped frame ${\hat{\bf X}}^w_t$ suffers from poor quality due to noisy flow estimation,  unexpected object occlusion, etc.  To improve the quality of warped frames, we propose to apply a {\it processing network} using ten residual blocks with embedded re-sampling to enlarge the receptive field, resulting in ${\hat{\bf X}}^{wp}_t$. Such methods have been used in denoising and deblurring applications to improve the quality of reconstruction.

Even with {\it processing network} included, we have observed that high frequency components are generally missing in ${\hat{\bf X}}^{wp}_t$. Motivated by~\cite{jo2018deep} that uses learned multi-frame residual information to improve the super resolution quality, we have attempted to apply the temporal recurrent network that is based on the ConvLSTM to capture and augment the high frequency priors to derive the $\tilde{\bf X}_t$ for temporal prediction, i.e.,
\begin{align}
  ({\bf c}_t, {\bf h}_t) = {\rm ConvLSTM}({\bf \Delta}_t, {\bf c}_{t-1}, {\bf h}_{t-1}), \label{eq:trn_priors}
\end{align} where ${\bf c}_t, {\bf h}_t$ is updated state  at $t$ slot with ${\bf c}_{t-1}$ used as a memory gate, and ${\bf h}_{t-1}$ as aggregated prior (i.e., probability, high frequency component, etc) update. ${\bf \Delta}_t$ is generally referred as the input feature vector. Here it is extracted features from reference frame $\hat{\bf X}_{t-1}$ in Fig.~\ref{sfig:inter_coding}.

\begin{figure}[t]
\centering
\subfigure[{KristenAndSara 1080p}]
{\includegraphics[scale=0.39]{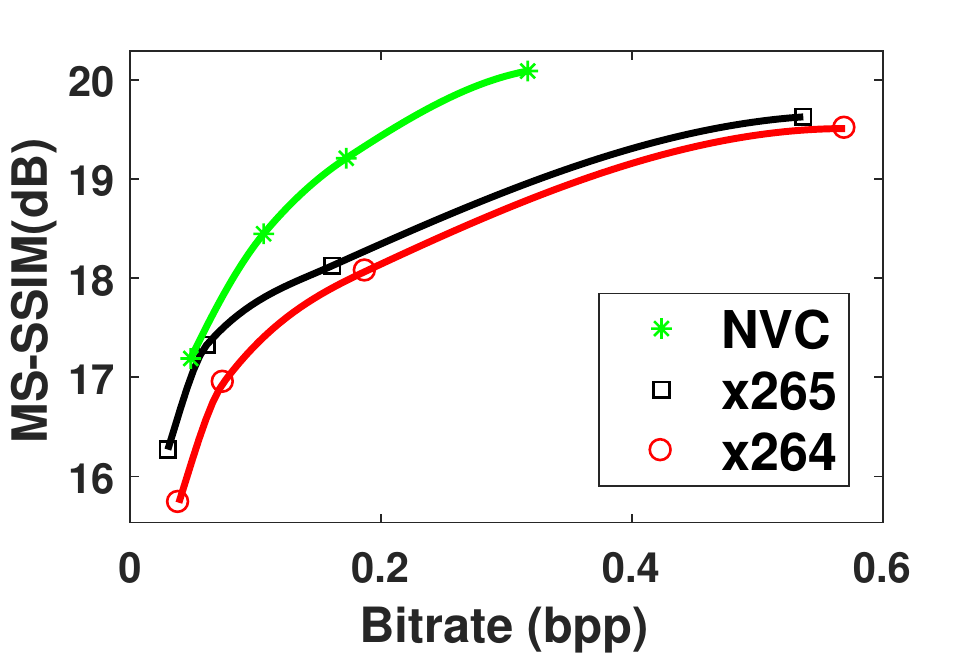}}
\subfigure[vidyo1 720p]{\includegraphics[scale=0.39]{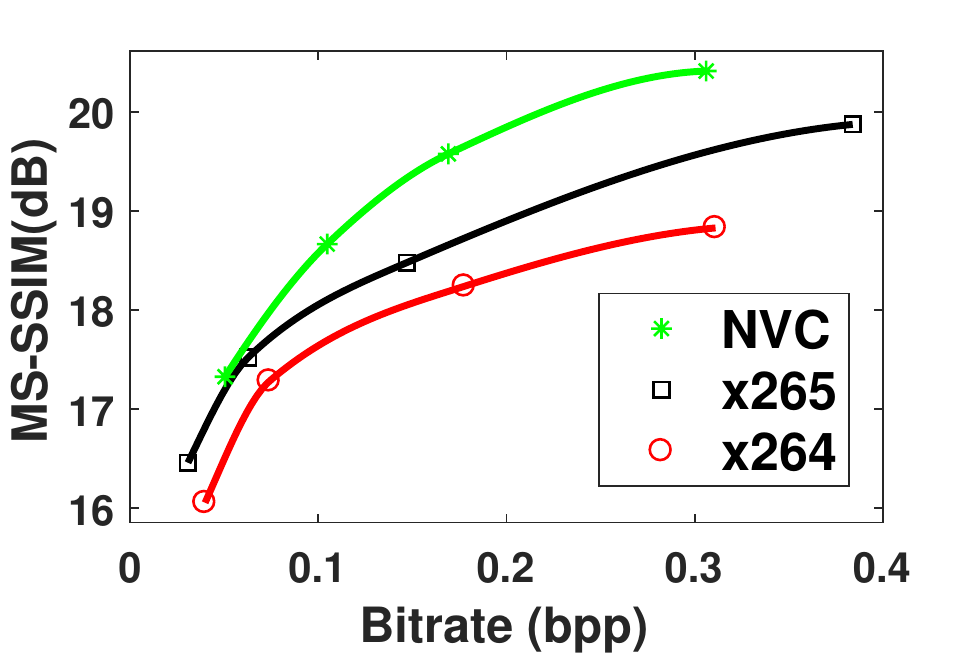}}
\subfigure[BasketBallDrive 1080p]{\includegraphics[scale=0.39]{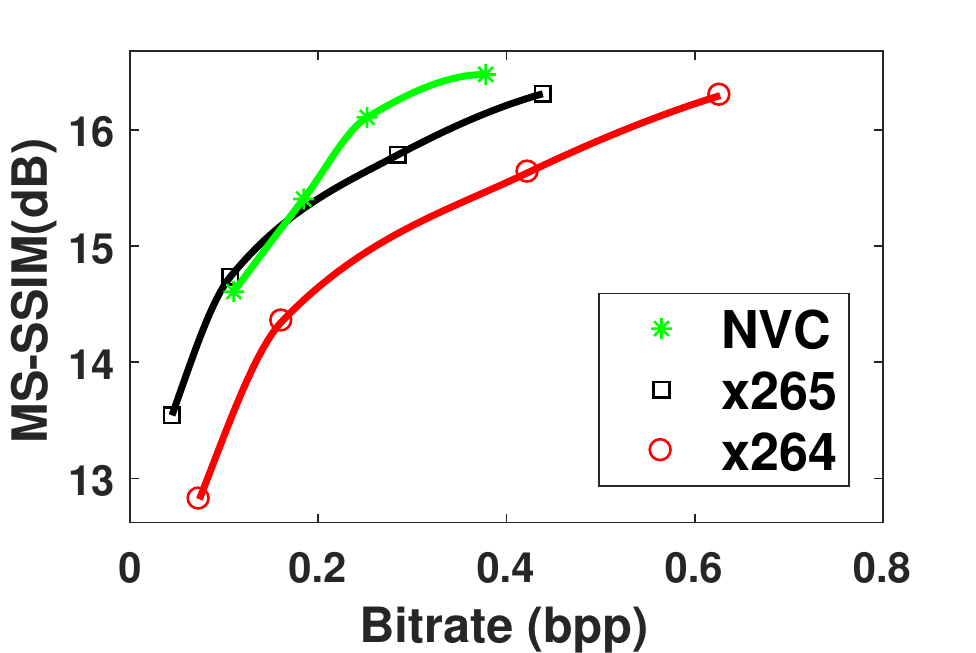}}
\subfigure[vidyo3 720p]{\includegraphics[scale=0.39]{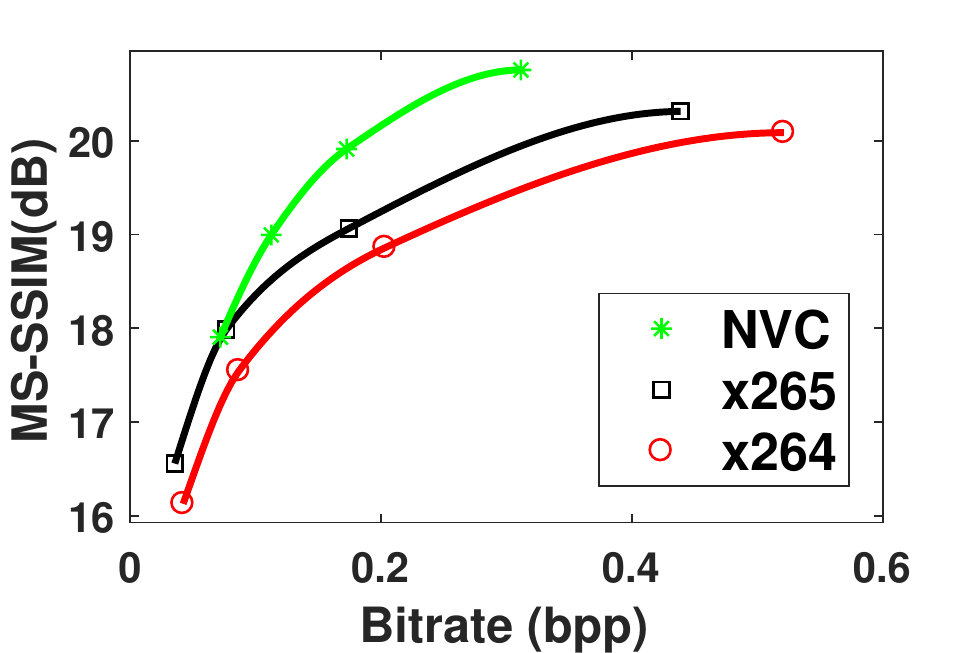}}
\subfigure[Cactus 1080p]{\includegraphics[scale=0.39]{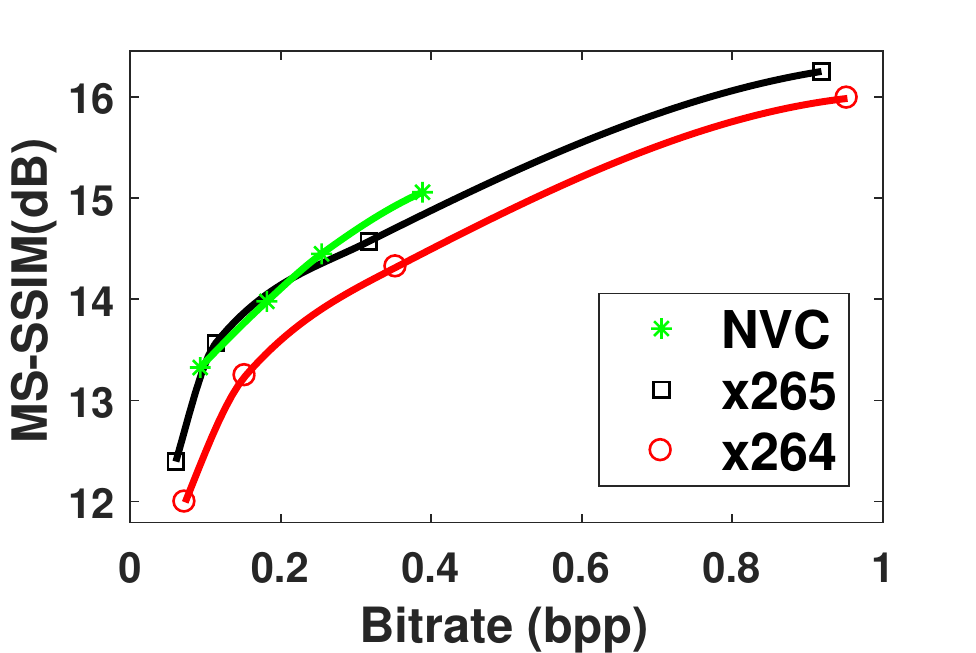}}
\subfigure[FourPeople 720p]{\includegraphics[scale=0.39]{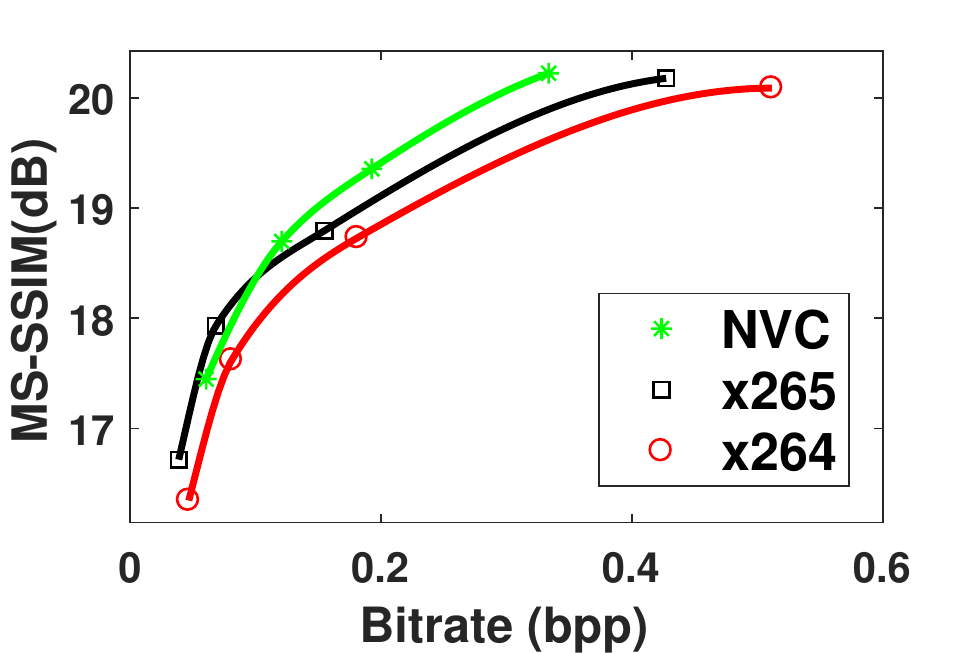}}
\caption{Illustration of the rate-distortion performance for six different sequences. Here we use $-10\log_{10}(1-d)$ to represent raw MS-SSIM ($d$) in dB scale.
}
\label{rd_curve}
\end{figure}

\subsection{Neural Residual Coding} \label{ssec:residual_coding}

Temporal residual coding shares the similar architecture as the intra coding shown in Fig.~\ref{sfig:intra_residual}, but with an augmented ConvLSTM to capture the aggregated temporal priors $\hat{\bf h}_t$ additionally for residual probability model improvement. Here, ${\bf \Delta}_t$ in Eq.~\eqref{eq:trn_priors} refers to the concatenated features in current frame $\hat{\bf Y}_t = {\bf CAT}(\hat{\bf X}_t, \rm{ICN}(\hat{\bf z}_t))$ in Fig.~\ref{sfig:intra_residual}.

We assume the same Gaussian distribution for residuals. Thus, Eq.~\eqref{eq:intra_feature} can be rewritten as
\begin{align}
  (\mu_r,\sigma_r) = {\prod_i} p(\hat{r}_i| \hat{r}_1,..., \hat{r}_{i-1}, \hat{\bf z}_t, {\bf h}_{t-1}), \label{eq:residual_probability}
\end{align} with aggregated temporal priors for residual probability modeling. $\hat{r}_i, i = 0, 1, 2, ...$ are pixels of residual frame $\hat{\bf r}_t$. We then can simply use the cumulative distribution function (CDF) to calculate the probability of each pixel:
\begin{align}
  &p_{\hat{r}|(\hat{r}_1,..., \hat{r}_{i-1},\hat{\bf z}_t, {\bf h}_{t-1})}(\hat{r}|\hat{r}_1,..., \hat{r}_{i-1},\hat{\bf z}_t, {\bf h}_{t-1})  \nonumber \\
  &\mbox{~~~~}= {\prod_i} (\mathcal{N}{(\mu_r,\sigma_r)} *\mathcal{U}(-\frac{1}{2},\frac{1}{2})) (\hat{r}_i).
  \label{eq:residual_dist}
 \end{align}

%When it comes to inter residual coding, a ConvLSTM in the decoder is added to transmit the temporal priors and updates its parameters using the following Eq.~\ref{Eq2}:
%\begin{equation}
%  h_t, c_t = ConvLSTM(y_t,h_{t-1},c_{t-1}),
%  \label{Eq2}
%\end{equation}

%Note that $h_t, c_t$ is updated state for next time step, $c_{t-1}$ is just used to update the ConvLSTM parameters and $h_{t-1}$ is used for both probablity model and ConvLSTM update. $y_t$ is the decoded feature in residual coding.
%Assuming the gain in temporal domain, we can further rewrite Eq.~\ref{Eq1} into Eq.~\ref{Eq3} to using extra temporal prior $h_{t-1}$ for better entropy modeling:
%\begin{equation}
%  \mu,\sigma = {\prod_i} p(x_i|x_1,...,x_{i-1},z_h,h_{t-1}),
%  \label{Eq3}
%\end{equation}
%We can simply use the cumulative distribution function (CDF) to calculate the probability of each symbol by Gaussian distribution:
%\begin{equation}
%  p_{\hat{y}|(\hat{z},h_{t-1})}(\hat{y}|(\hat{z},h_{t-1})) = {\prod_i} (\mathcal{N(\mu,\sigma)} *\mathcal{U}(-\frac{1}{2},\frac{1}{2})) (\hat{y}_i),
 % \label{Eq4}
 %\end{equation}

%Overall,intra coding is treated as a relatively independent module in our framework. For inter residual coding, we proposed a ConvLSTM in the decoder to capture temporal dependence which is intended to achieve spatio-temporal prior fusion to make more accurate probability prediction. Note that the inter residual coding network shares weights in the time domain.

\section{End-to-End Training Strategy}
\label{sec:end_to_end_training}

It is difficult to train multiple networks jointly on-the-fly for our NVC. Thus, we pre-train the intra coding and flow coding networks first, followed by the jointly training with pre-trained nets for an overall rate-distortion optimization~\cite{Gary_RDO}, i.e.,
\begin{align}
  L =  {\frac{\lambda_1}{n}{\sum_{t=0}^{n}}{\mathbb D}_1(\hat{{\bf X}}_t,{{\bf X}_t})}
    & + {\frac{\lambda_2}{n}{\sum_{t=0}^{n}}{\mathbb D}_2(\hat{\bf X}_t^{wp},{{\bf X}_t})} \nonumber\\
    & +  R_s+{\frac{1}{n-1}{\sum_{t=1}^{n}}R_t}, \label{eq:rdo}
\end{align} where ${\mathbb D}_1$ is measured using MS-SSIM, and ${\mathbb D}_2$ is the warp loss evaluated using $L$1 norm and total variation loss. $R_s$ represents the bit rate of intra frame and $R_t$ is the bit rate of inter frames including bits for residual and flow. Currently, $\lambda_1$ and $\lambda_2$ will be adapted according to the specified overall bit rate and bit consumption percentage of flow information in inter coding.

\begin{table}[t]
  \centering
  \caption{BD-Rate Gain of our NVC versus HEVC MP}
  \label{tab:BDRate}
  \begin{tabular}{|c|c|c|c|}
    \hline
       Sequence & Resolution & FPS &  BD-Rate Gain \\
    \hline
     KristenAndSara & 1920x1080 &  60 & -96.96\%\\
     \hline
     vidyo1 & 1280x720 & 60 & -54.32\%\\
    \hline
     BasketBallDrive & 1920x1080 & 50 & -12.07\%  \\
    \hline
     vidyo3 & 1280x720 & 60 & -49.15\% \\
    \hline
     Cactus & 1920x1080 & 50&  -2.27\%\\
     \hline
     FourPeople & 1280x720 & 60&  -13.92\%\\
    \hline
    Ave. & & & -38.12\%\\
    \hline
  \end{tabular}
\end{table}

We start at a learning rate (LR) of $10^{-4}$ and reduce it by half every 30 epochs.
We choose vimeo dataset~\cite{xue2017video}  and randomly crop the data into 192{$\times$}192 spatially, as our training set. To well balance the efficiency of temporal information learning and training memory consumption, we have enrolled 5 frames to train the video compression framework and shared the weights for the rest in the time domain.

\section{Experimental Studies}
We have evaluated our NVC framework using BD-Rate but with distortion measured using MS-SSIM in dB scale. All simulation candidates, i.e., NVC, H.264/AVC HP, HEVC MP, use IPPP mode to encode the video data with the same GOP of 8 for fair comparison.
Well recognized x264 and x265 softwares are used for H.264/AVC HP and HEVC MP, respectively.
Several standard test sequences in different content classes are tested, and results have shown that our NVC presents a noticeable BD-Rate improvement compared with traditional H.264/AVC HP and HEVC MP as shown in Fig.~\ref{rd_curve}. In the meantime, BD-Rate reduction compared with HEVC MP is calculated
 and shown in Table~\ref{tab:BDRate}, where $\approx$ 38.12\% BD-Rate gain is reported on average.

\begin{figure}[t]
     \centering
     \includegraphics[scale=0.23]{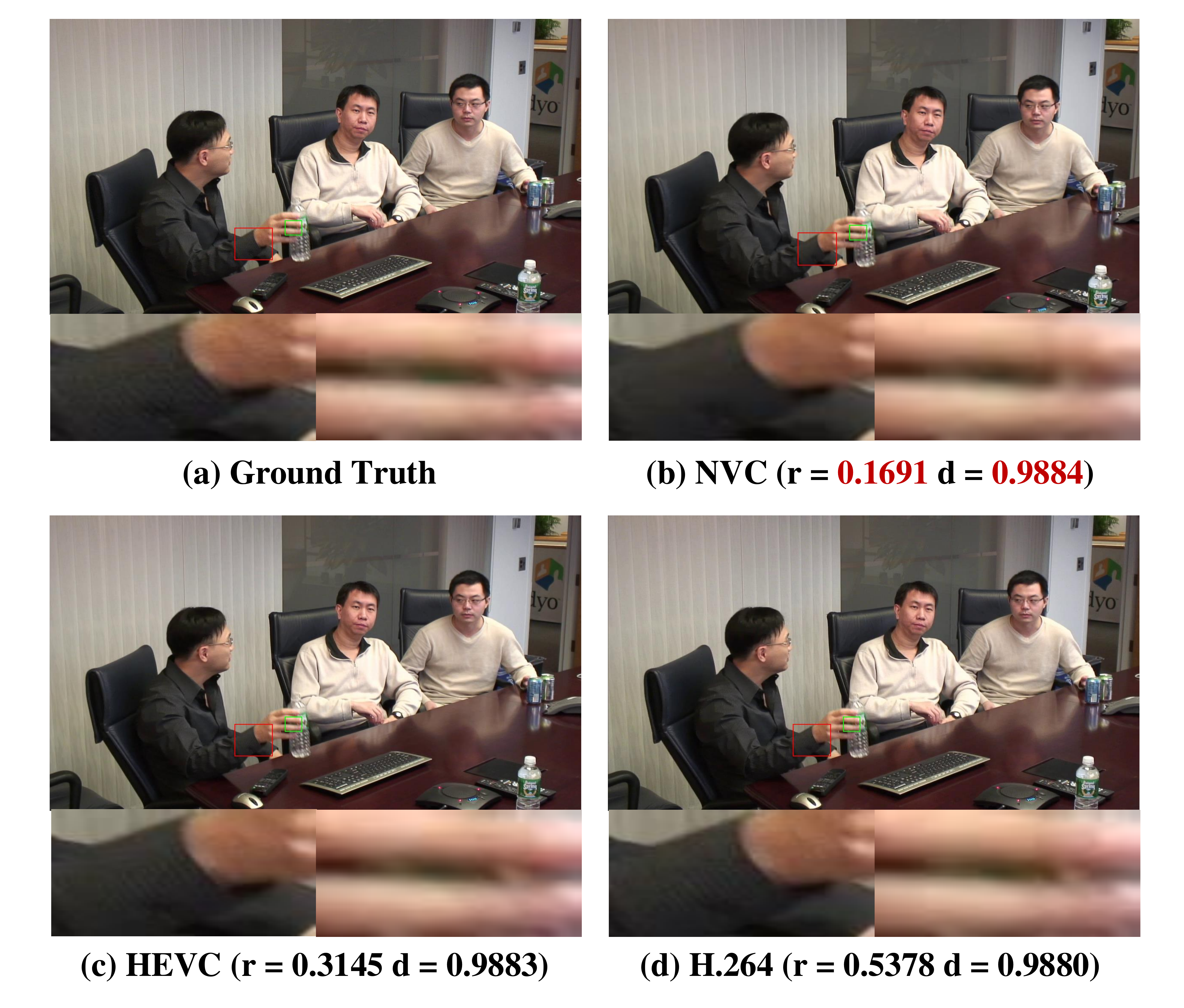}
     \caption{Snapshots of performance evaluation  at the same quality ($r$ for bit rate at bits per pixel, and $d$ for MS-SSIM).}
     \label{subjective_test}
\end{figure}

 %using \cite{bjontegaard2001calculation}. Note that the distortion is measured by MS-SSIM as shown in Table~\ref{tab:BDRate}.
We have also provided snapshots for original raw, NVC, H.264/AVC HP and HEVC MP encoded frames, respectively in Fig.~\ref{subjective_test}. At the similar quality with MS-SSIM close to 0.988, our NVC has demonstrated significant bit rate reduction compared with the H.264/AVC HP and HEVC MP.

%show our subjective result using adjacent MS-SSIM and find our method can save lots of bit rates compared with H.264 and HEVC in Fig.~\ref{subjective_test1}.

\section{Conclusion}
\label{sec:print}
We proposed a neural video compression framework leveraging the spatio-temporal priors jointly, which outperforms the well-known H.264/AVC and HEVC with noticeable BD-Rate gains (i.e., 38\% on average), resulting in the state-of-the-art coding performance. Spatial priors are derived from the downscaled image representations and temporal priors are captured using a recurrent network. As for future studies, bidirectional temporal prediction is one of the primary focuses to further improve the compression efficiency. In the meantime, adaptive bit rate allocation among intra texture, flow motion and residual with optimal rate-distortion efficiency will be another exploration avenue.

\bibliographystyle{IEEEbib}
\bibliography{video_compression_v6}

\begin{thebibliography}{10}

\bibitem{HEVC}
G.~J. Sullivan, J.~Ohm, W.~Han, and T.~Wiegand,
\newblock ``Overview of the high efficiency video coding (hevc) standard,''
\newblock {\em IEEE Transactions on Circuits and Systems for Video Technology},
  vol. 22, no. 12, pp. 1649--1668, Dec 2012.

\bibitem{VVC_spec}
B.~Bross, J.~Chen, and S.~Liu,
\newblock ``Versatile video coding (draft 3),''
\newblock {\em Doc. JVET L1001}, Oct. 2018.

\bibitem{Gary_VideoCompressionBasics}
G.~J. Sullivan and T.~Wiegand,
\newblock ``Video compression - from concepts to the h.264/avc standard,''
\newblock {\em Proceedings of the IEEE}, vol. 93, no. 1, pp. 18--31, Jan 2005.

\bibitem{balle2018variational}
Johannes Ball{\'e}, David Minnen, Saurabh Singh, Sung~Jin Hwang, and Nick
  Johnston,
\newblock ``Variational image compression with a scale hyperprior,''
\newblock {\em arXiv preprint arXiv:1802.01436}, 2018.

\bibitem{Liu_2018_CVPR_Workshops}
Haojie Liu, Tong Chen, Qiu Shen, Tao Yue, and Zhan Ma,
\newblock ``Deep image compression via end-to-end learning,''
\newblock in {\em The IEEE Conference on Computer Vision and Pattern
  Recognition (CVPR) Workshops}, June 2018.

\bibitem{rippel2017real}
Oren Rippel and Lubomir Bourdev,
\newblock ``Real-time adaptive image compression,''
\newblock {\em arXiv preprint arXiv:1705.05823}, 2017.

\bibitem{mentzer2018conditional}
Fabian Mentzer, Eirikur Agustsson, Michael Tschannen, Radu Timofte, and Luc
  Van~Gool,
\newblock ``Conditional probability models for deep image compression,''
\newblock in {\em IEEE Conference on Computer Vision and Pattern Recognition
  (CVPR)}, 2018, vol.~1, p.~3.

\bibitem{chen2017deepcoder}
Tong Chen, Haojie Liu, Qiu Shen, Tao Yue, Xun Cao, and Zhan Ma,
\newblock ``Deepcoder: A deep neural network based video compression,''
\newblock in {\em Visual Communications and Image Processing (VCIP), 2017
  IEEE}. IEEE, 2017, pp. 1--4.

\bibitem{lu2018dvc}
Guo Lu, Wanli Ouyang, Dong Xu, Xiaoyun Zhang, Chunlei Cai, and Zhiyong Gao,
\newblock ``Dvc: An end-to-end deep video compression framework,''
\newblock {\em arXiv preprint arXiv:1812.00101}, 2018.

\bibitem{rippel2018learned}
Oren Rippel, Sanjay Nair, Carissa Lew, Steve Branson, Alexander~G Anderson, and
  Lubomir Bourdev,
\newblock ``Learned video compression,''
\newblock {\em arXiv preprint arXiv:1811.06981}, 2018.

\bibitem{gers2001long}
Felix Gers,
\newblock {\em Long short-term memory in recurrent neural networks},
\newblock Ph.D. thesis, Verlag nicht ermittelbar, 2001.

\bibitem{xingjian2015convolutional}
SHI Xingjian, Zhourong Chen, Hao Wang, Dit-Yan Yeung, Wai-Kin Wong, and
  Wang-chun Woo,
\newblock ``Convolutional lstm network: A machine learning approach for
  precipitation nowcasting,''
\newblock in {\em Advances in neural information processing systems}, 2015, pp.
  802--810.

\bibitem{H264AVC}
T.~Wiegand, G.~J. Sullivan, G.~Bjontegaard, and A.~Luthra,
\newblock ``Overview of the h.264/avc video coding standard,''
\newblock {\em IEEE Transactions on Circuits and Systems for Video Technology},
  vol. 13, no. 7, pp. 560--576, July 2003.

\bibitem{bjontegaard2001calculation}
Gisle Bjontegaard,
\newblock ``Calculation of average psnr differences between rd-curves,''
\newblock {\em VCEG-M33}, 2001.

\bibitem{wang2003multiscale}
Zhou Wang, Eero~P Simoncelli, and Alan~C Bovik,
\newblock ``Multiscale structural similarity for image quality assessment,''
\newblock in {\em The Thrity-Seventh Asilomar Conference on Signals, Systems \&
  Computers, 2003}. Ieee, 2003, vol.~2, pp. 1398--1402.

\bibitem{he2016deep}
Kaiming He, Xiangyu Zhang, Shaoqing Ren, and Jian Sun,
\newblock ``Deep residual learning for image recognition,''
\newblock in {\em Proceedings of the IEEE conference on computer vision and
  pattern recognition}, 2016, pp. 770--778.

\bibitem{Gary_RDO}
G.~J. Sullivan and T.~Wiegand,
\newblock ``Rate-distortion optimization for video compression,''
\newblock {\em IEEE Signal Processing Magazine}, vol. 15, no. 6, pp. 74--90,
  Nov 1998.

\bibitem{ilg2017flownet}
Eddy Ilg, Nikolaus Mayer, Tonmoy Saikia, Margret Keuper, Alexey Dosovitskiy,
  and Thomas Brox,
\newblock ``Flownet 2.0: Evolution of optical flow estimation with deep
  networks,''
\newblock in {\em 2017 IEEE conference on computer vision and pattern
  recognition (CVPR)}. IEEE, 2017, pp. 1647--1655.

\bibitem{jo2018deep}
Younghyun Jo, Seoung Wug~Oh, Jaeyeon Kang, and Seon Joo~Kim,
\newblock ``Deep video super-resolution network using dynamic upsampling
  filters without explicit motion compensation,''
\newblock in {\em Proceedings of the IEEE Conference on Computer Vision and
  Pattern Recognition}, 2018, pp. 3224--3232.

\bibitem{xue2017video}
Tianfan Xue, Baian Chen, Jiajun Wu, Donglai Wei, and William~T Freeman,
\newblock ``Video enhancement with task-oriented flow,''
\newblock {\em arXiv preprint arXiv:1711.09078}, 2017.

\end{thebibliography}

\end{document}